\documentclass[a4paper,11pt]{article}
\usepackage{jinstpub} 
\usepackage{upgreek}

\title{\boldmath Characterization of MOSS for the ALICE ITS3 for the LHC Run 4}

\author[1]{L. Terlizzi\note{Corresponding author.}}
\author{, on behalf of the ALICE Collaboration}
\affiliation[1]{University of Turin,\\
1, Via Pietro Giuria, Turin, Italy}

\emailAdd{livia.terlizzi@unito.it}

\abstract{ALICE (A Large Ion Collider Experiment) is one of the four main experiments at the CERN Large Hadron Collider (LHC), and it is mainly designed to study heavy-ion collisions at ultra-relativistic energies. In view of the LHC Run 4, foreseen to start in 2030, ALICE will replace the three innermost cylindrical layers of its current inner tracking system (ITS2) during the Long Shutdown 3 (2026-2029). The new system, ITS3, will improve the pointing resolution by a factor of two over a large momentum range and the tracking efficiency at low transverse momenta of \textit{p}$_\textrm{T}$ < 0.3
GeV/\textit{c}. It will consist of ultra-thin, $\leq$ 50 $\upmu$m, stitched wafer-scale Monolithic Active Pixel Sensors (MAPS), built using the 65 nm CMOS imaging process. These sensors are flexible and can be bent to form a cylindrical barrel with a radial distance from the beam pipe as low as 19 mm for the innermost layer and a very low material budget of 0.09 X/X$_0$ on average per layer. The development of the ITS3 includes a number of cutting-edge R$\&$D efforts: the production and characterization of the MAPS in the 65 nm CMOS process, the fabrication of the stitched wafer-scale MAPS, and the development of an ultra-light detector mechanics and an air cooling system. In particular, the 65 nm CMOS technology for particle tracking and radiation hardness was validated with a set of test structures called Multi-Layer Reticle 1 (MLR1). In mid 2023 the first stitched prototypes called MOnolithic Stitched Sensors (MOSS) have been produced with the primary goals
of demonstrating the feasibility of the stitching process and of studying the yield and performance of wafer-scale sensors, in view of the production of the ITS3 final-size full-functionality prototype sensor chip.
A single MOSS chip measures 1.4 $\times$ 25.9 cm$^2$ and has a total of 6.7 million pixels. It is composed of one left endcap, 10 repeated sensor units with 8 pixel matrices each, to increase power granularity and hence resilience to manufacturing faults: 256 $\times$ 256 pixels with 22.5 $\upmu$m pitch in each top matrix and 320 $\times$ 320 pixels with 18 $\upmu$m pitch in each bottom matrix. The different layouts are used to compare the yield depending on the densities and spacing margins.
This contribution focuses on the results from the characterisation campaign of the stitched sensors in the laboratory and in beam tests, including the verification of power domain impedances, Digital-to-Analog Converter (DAC) performance, pixel front-end readout response, threshold and fake-hit rate scans. Test results have proven that MOSS has an efficiency higher than 99$\%$ with a fake hit rate lower than 0.1 pixel$^{-1}$ s$^{-1}$, which satisfies ITS3 sensor requirements for LHC Run 4.}


\arxivnumber{} 

\begin{document}
\maketitle
\flushbottom

\section{Introduction}
\label{sec:intro}

During the LHC Long Shutdown 3 (LS3), ALICE aims at replacing the three innermost layers of the current Inner Tracking System (ITS2)~\cite{a} with a detector featuring wafer-scale, bent monolithic CMOS sensors ~\cite{b,c}. ITS3 will significantly enhance tracking and vertexing performance, doubling the pointing resolution across the entire transverse momentum range and improving tracking efficiency by up to 40$\%$ at low \textit{p}$_\textrm{T}$ compared to ITS2. It features true cylindrical layers, achieved using wafer-scale large area sensors fabricated with stitching technology, measuring 266 $\times$ 98 mm$^2$. These ultra-thin silicon sensors, just 50 $\upmu$m thick, will be placed at 19 mm for the innermost layer, compared to the 22 mm of the ITS2, which features a non-bent design. The bent silicon die provides a stable, self-supporting structure, strongly reducing the need for mechanical support components. The silicon sensors will be the only active components, reducing the average material budget from 0.35$\%$ X$_0$ to 0.09$\%$ X$_0$ per layer. Air cooling will replace the current water system, with power dissipation kept below 40 mW/cm$^2$. The ITS3 development involves advanced R$\&$D efforts, including MAPS production with 65 nm TPSCo  (Tower Partners Semiconductor Co.) CMOS imaging technology ~\cite{d}, stitched wafer-scale sensors, and ultra-light air-cooled design. A key milestone was validating the 65 nm CMOS process for particle tracking and radiation hardness through the Multi-Layer Reticle 1 (MLR1) test structures~\cite{e, f}. MLR1 aimed to build expertise, confirm the technology’s suitability for particle detection, and evaluate prototype circuits and pixel sensors. The second silicon submission, Engineering Run 1 (ER1), aims to demonstrate the feasibility of the stitching process, assess wafer-scale sensor yield, and gain expertise. The stitching technique~\cite{g} allows the production of larger devices than the typical 3 $\times$ 2 cm$^2$ design reticle by dividing the reticle into sub-frames that are precisely aligned and exposed onto the wafer. Peripheral structures like outer edges and corners are placed in dedicated sub-frames, enabling the creation of large chips approaching the wafer's diameter. In mid-2023 the first stitched prototypes, MOnolithic Stitched Sensors (MOSS), were produced. This contribution summarizes results from laboratory and tests beam, covering power domain impedance, DAC scans, pixel readout response, threshold and fake-hit rate scans, and radiation hardness validation. The expected ITS3 levels are below 10$^{13}$ 1 MeV n$_{\textrm{eq}}$ cm$^{-2}$ and 1 Mrad, with detection efficiency above 99$\%$ and fake hit rate below 0.1 pixel$^{-1}$ s$^{-1}$, corresponding to $\sim$ 2 $\times$ 10$^{-6}$ pixel$^{-1}$ event$^{-1}$ at a 50 kHz readout rate, within ITS3 requirements for LHC Run 4 ~\cite{b}).




\section{The MOnolithic Stitched Sensor (MOSS)}

\begin{figure}[htbp]
\centering
\includegraphics[width=.7\textwidth]{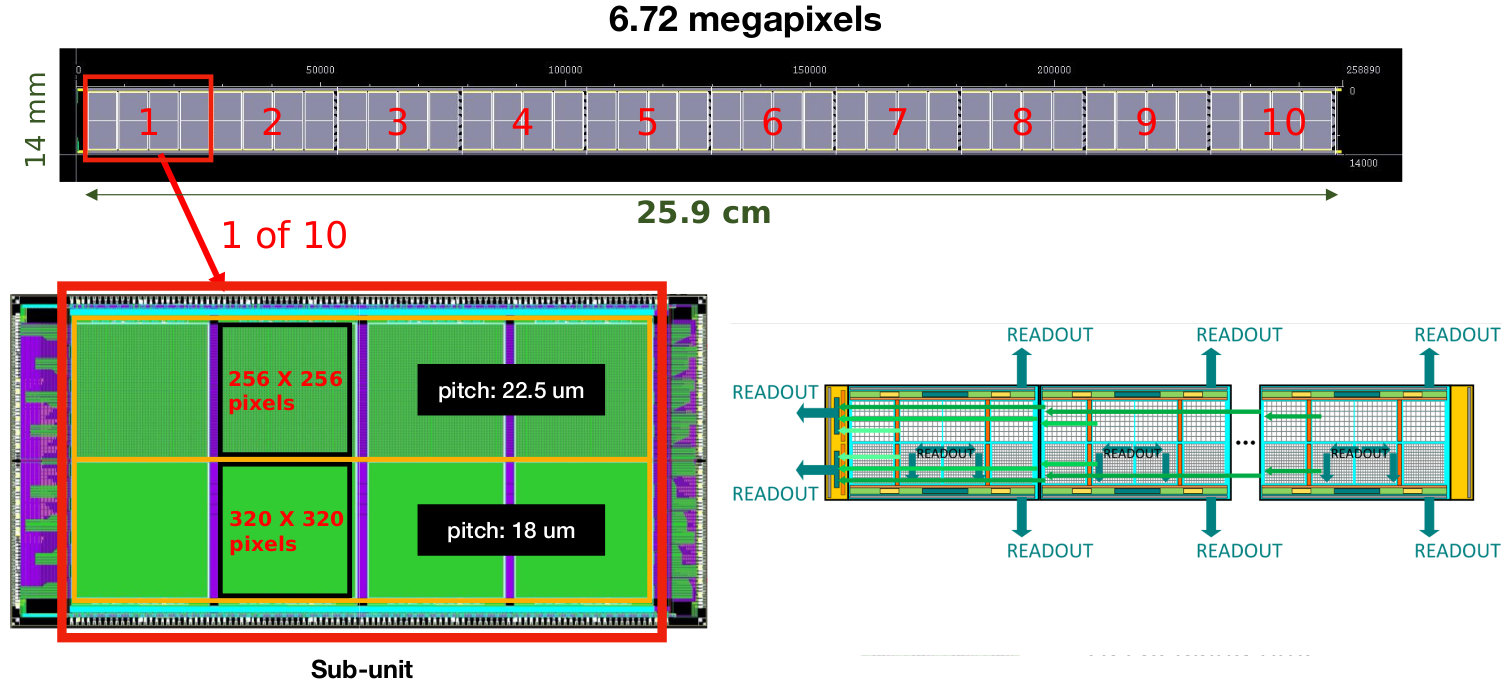}
\caption{Structure of the MOSS sensor. The top section shows the full sensor composed of ten stitched Repeated Sensor Units (RSUs) and two end-cap regions. The bottom sections zoom into a half-RSU unit (left), illustrating the pixel matrix and layout, and the readout architecture (right), highlighting data and control routing through the Stitched Communication Backbones.\label{fig:moss}}
\end{figure}

The MOSS chip, shown in figure~\ref{fig:moss},  aims to validate large-area stitched sensor integration, low-mass design, and compliance with performance targets such as high detection efficiency and low fake-hit rate. It consists of ten Repeated Sensor Units (RSUs) arranged using stitching, enclosed by two smaller end-cap regions. The overall dimensions of the chip are 259 $\times$ 14 mm$^2$, with 6 MOSS chips produced per wafer. Each RSU is divided into two half-units, for a total of 20 half-units per chip, featuring different pixel array pitches: top half-unit has 4 matrices, also referred to as regions, of 256 $\times$ 256 pixels with a 22.5 $\upmu$m pitch; bottom half-unit has 4 regions of 320 $\times$ 320 pixels with an 18.0 $\upmu$m pitch. The wafer also includes several babyMOSS chips, each containing a single RSU, suitable for irradiation purposes. The different circuit densities in the top and bottom half-units allow for design optimization, with the bottom half-unit having a denser layout and the top half-unit focusing on wider spacing of the interconnecting metal structures. This design choice is driven by yield studies, which indicate that reducing the density of interconnects in the top half-unit helps mitigate potential defects and improves fabrication yield, ensuring better manufacturability without compromising ITS3 performance. The sensor contains 6.72 million pixels, with each half-unit independently operable, featuring its own power, peripheral circuits, and I/O interfaces. This design, informed by power distribution constraints and yield studies, enhances reliability by preventing failures in one half-unit from affecting the other. The total analog power density is 7 mW/cm$^2$ for the top half-unit and 11 mW/cm$^2$ for the bottom half-unit. Metal interconnections across stitching boundaries ensure seamless integration with the power distribution from the left to the right end-cap. Two Stitched Communication Backbones facilitate signaling between RSUs and the peripheral readout circuitry located in the left end-cap region of the chip, which hosts the control interface and multiple point-to-point data readout lines. This backbone enables full control and readout via interfaces on the left end-cap, supported by 20 distributed ports along the chip edges, each serving an individual half-unit. \\
An intense characterization campaign has been performed on MOSS sensor, via mass testing in CERN laboratory and beam tests. Results from both are presented in the next sections.

\section{Mass testing at CERN}

\begin{figure}[htbp]
\centering
\includegraphics[width=.45\textwidth]{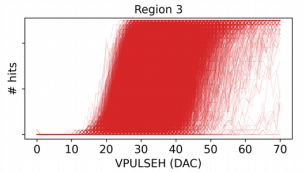}
\caption{Example S-curves for all pixels in one region of a top half-unit from a representative wafer. \label{fig:scurve}}
\end{figure}

\begin{figure}[htbp]
\centering
\includegraphics[width=.72\textwidth]{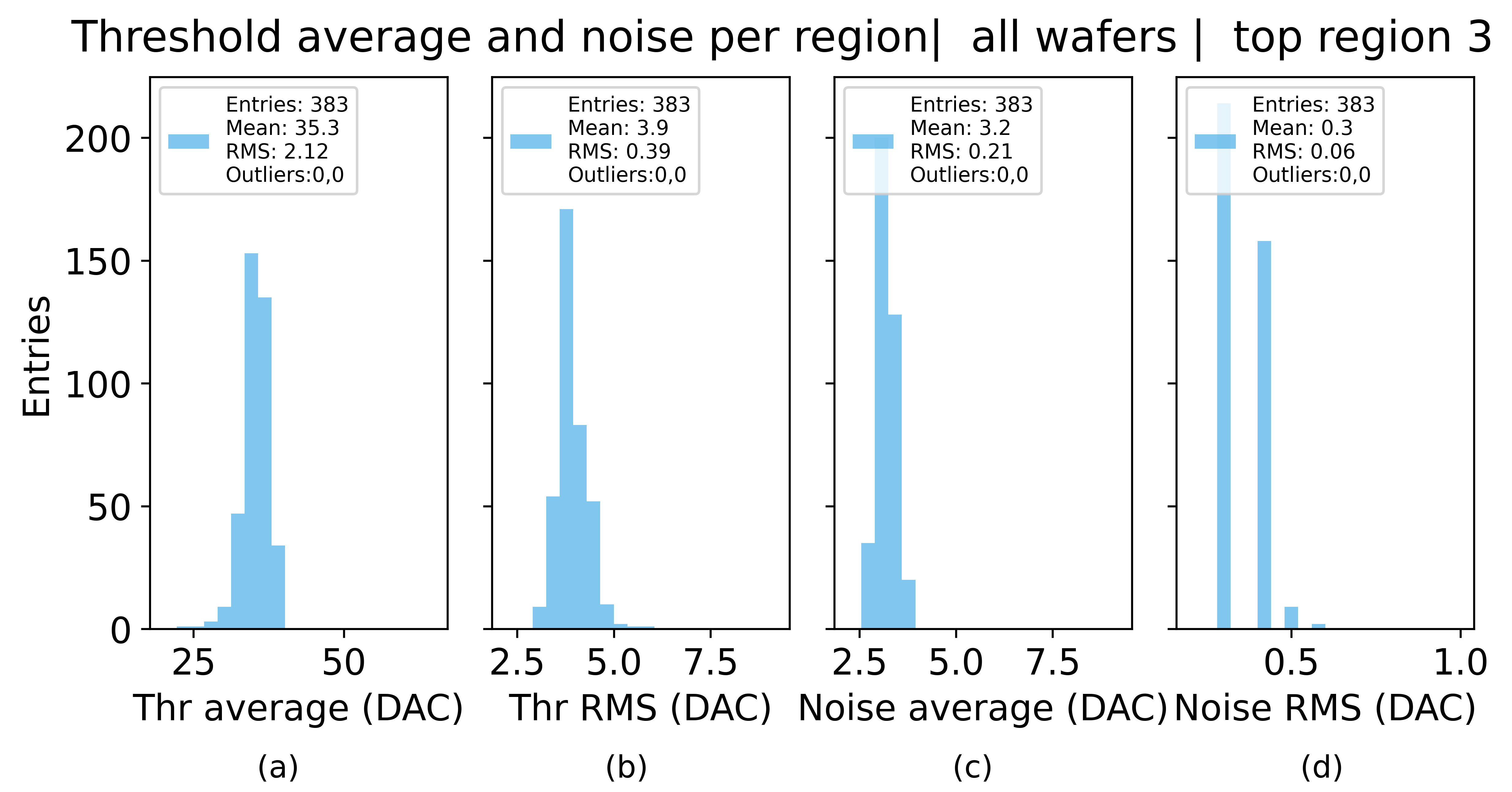}
\caption{Summary of average threshold (a) and noise (c) values for region 3 of the top half-unit, aggregated over all tested wafers. The RMS values of threshold (b) and noise (d) represent the spread across all region 3 top half-units from all wafers, characterizing the inter-region variation. The observed jumps in RMS reflect wafer-to-wafer or layout-induced process variations. \label{fig:thr}}
\end{figure}

A custom setup was developed for systematic mass testing of MOSS sensors, providing bias voltages and control signals. Yield analysis was conducted at multiple granular levels, from half units down to individual regions. A -1.2 V bias can be applied to the substrate to improve charge collection efficiency. In total, 82 MOSS chips were bonded on carrier boards, of which 46 chips (equivalent to 3600 regions) were tested without applying reverse bias. Testing with a reverse bias of -1.2 V is currently ongoing. This section focuses exclusively on results obtained without reverse bias. \\
Initial testing involved independently powering the 20 half units to assess yield and performance variations based on circuit density. Impedance measurements were conducted between all eight power nets for each half unit, resulting in 28 net pair combinations per half unit. These measurements are crucial for identifying potential faults before power ramping the chip. Subsequently, power ramping was performed to bring each half unit to its nominal voltage, carefully monitoring for any (transient) high currents during the process. Systematic measurement with a thermal camera for each half unit are conducted to detect shorts between metal layers. Impedance measurement is repeated after power ramping to eventually confirm opening of the shorts. Half Units that successfully pass the power ramp are functionally tested. 
The functional testing begins with a register scan, where all 7680 applicable registers in each chip are tested using predefined bit patterns to ensure proper functionality. This is followed by a DAC scan to evaluate the linearity of the in-chip DACs. MOSS includes features such as analog charge injection for testing the front-end electronics and digital pulses for verifying pixel logic functionality. Digital and analogue scans involve 25 repeated injections per pixel to assess their response. Dead pixels are those that do not respond at all. Noisy pixels fire inconsistently, registering hits in fewer than 25 times. Inefficient pixels exhibit partial responsiveness, recording a hit count between 1 and 24 instead of the expected 25. \\
In the threshold scan the injected charge, $Q_{\textrm{inj}}=C_{\textrm{inj}}$ $\times$ $V_{\textrm{PULSEH}}$, is varied by scanning $V_{\textrm{PULSEH}}$ from 0 to 70 DAC in steps of 1. For each step, 25 charge pulses are injected into every pixel. As the injected charge approached the pixel's threshold, the number of hits per step increased, eventually reaching a plateau, producing the so-called "S-curve". Figure~\ref{fig:scurve} shows an example of S-curves obtained from all pixels in a single region of one top half-unit on a representative wafer. The derivative of the S-curve yields a Gaussian distribution, where the mean corresponds to the pixel threshold and the standard deviation represents the equivalent noise charge (ENC). Figure~\ref{fig:thr} presents aggregated results for region 3 of the top half-unit across all tested wafers, including the average and RMS values of the threshold and noise distributions. During the analysis, noisy pixels, i.e. those that fired without injection, and bad pixels, i.e. those that failed to fire when injected, are identified. Threshold scans are conducted on each region of each half unit, producing S-curves for every pixel. This approach provided a comprehensive evaluation of pixel thresholds and noise characteristics, enabling precise performance assessment. \\
The fake-hit rate scan involves reading out multiple chip events to identify noisy pixels. Each region of the chip is read out 100000 times without any external stimulus or internal pulsing, allowing the detection of pixels that fire above the threshold. For each region, the number of hits is recorded and used to calculate the fake-hit rate, defined as the number of hits per pixel per second in the absence of external stimuli. Pixels that fire more frequently than a predefined threshold (hits/trigger > 10$^{-2}$) are classified as noisy and subsequently masked. The number of masked pixels per region is recorded, and the fake-hit rate is evaluated again after masking. On average, it is sufficient to mask 1-2 pixels per region to get a fake-hit rate within the ITS3 requirements, i.e. less than 10$^{-6}$ hits/pixel/event ~\cite{b}. \\
In table~\ref{tab:i} the MOSS scans success rate is shown. The yield loss is cumulative, meaning that failures in earlier scans carry over to subsequent ones, progressively reducing the overall success rate. The results demonstrate consistent and homogeneous performance, indicating a high level of reliability with a final yield higher than 94$\%$, indicating a high level of reliability.

\begin{table}[htbp]
\centering
\caption{MOSS scans success rates.\label{tab:i}}
\smallskip
\begin{tabular}{cccccc}
\hline
Register scan&DAC scan& Digital Scan & Analogue scan & Threshold scan & Fake-hit rate scan \\
\hline
99.85$\%$ & 99.60 $\%$ & 98.6 $\%$ & 99.7 $\%$ & 98 $\%$ & 94.7 $\%$ \\
\hline
\end{tabular}
\end{table}

\section{Test beams characterization}

\begin{figure}[htbp]
\centering
\includegraphics[width=.8\textwidth]{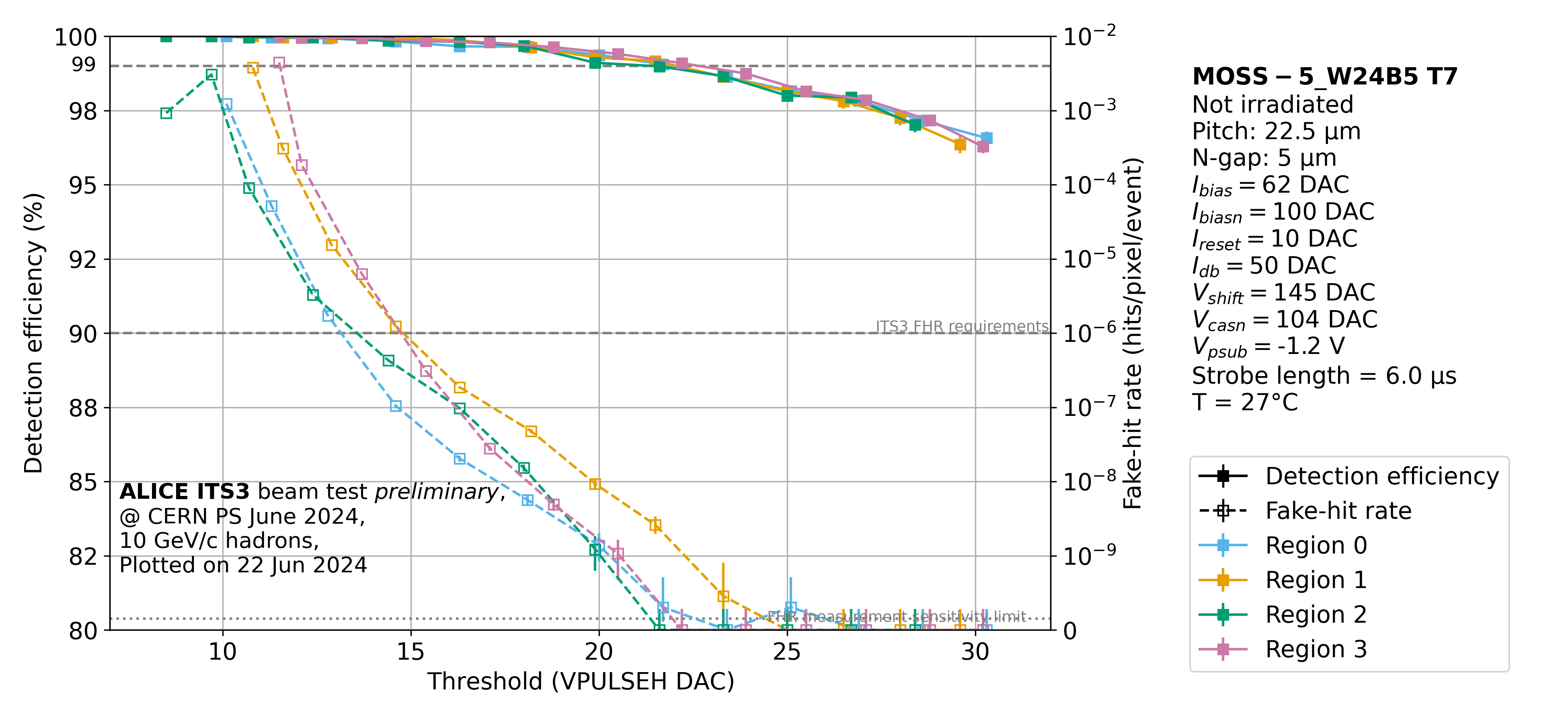}
\qquad
\includegraphics[width=.8\textwidth]{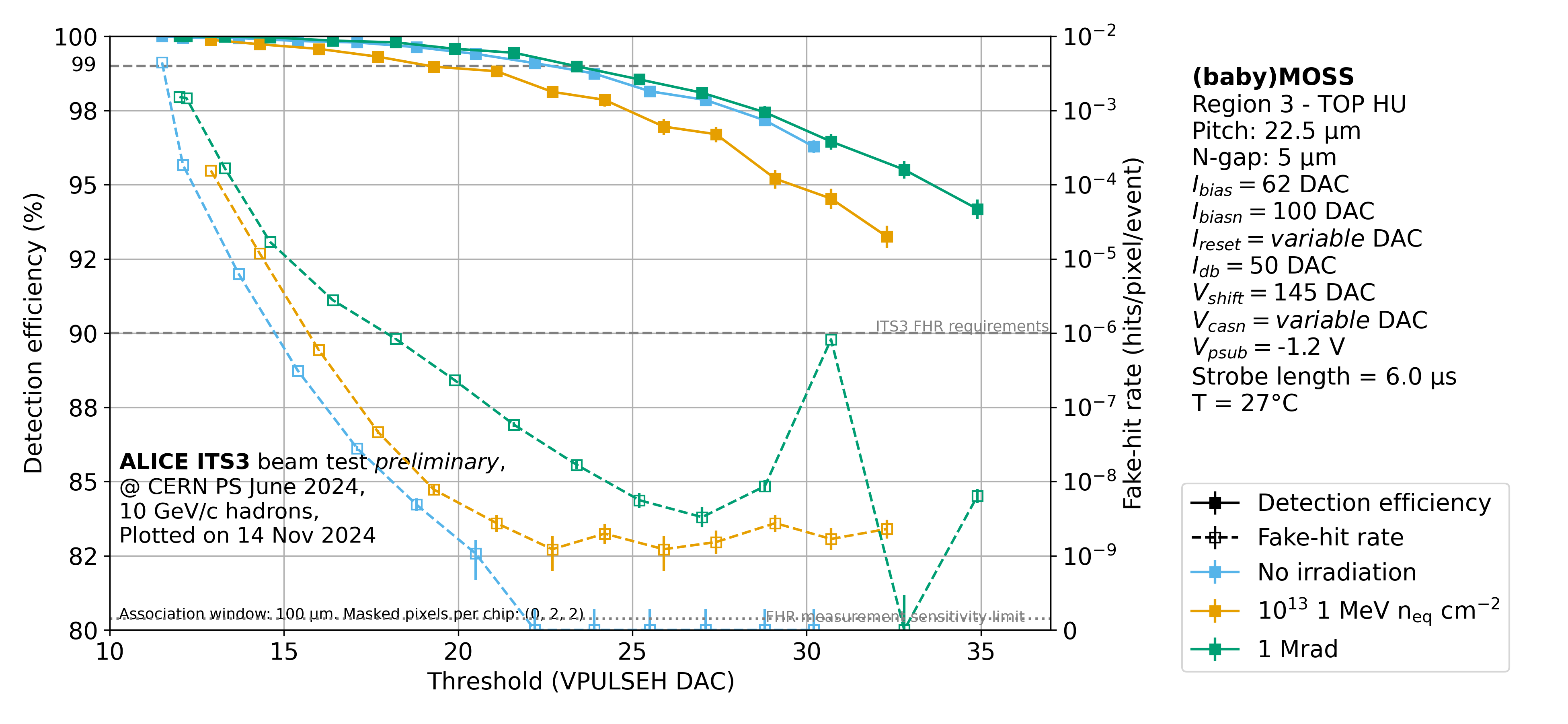}
\caption{Detection efficiency (solid lines) and fake-hit rate (dashed lines) as a function of threshold, measured with 10 GeV/\textit{c} $\uppi^{-}$. Top: four regions of a top half-unit in a non-irradiated MOSS sensor. Bottom: one region of a top half unit, comparing results across various irradiation levels. \label{fig:tb}}
\end{figure}

The goal of beam tests is to evaluate detection efficiency and spatial resolution as a function of threshold, before and after TID (Total Ionizing Dose) and NIEL (Non-Ionizing Energy Loss) irradiation. These tests were conducted at the CERN PS beam-test facility (beamline T10) with 10 GeV $\uppi^{-}$ beams. A total of 15 chips were studied, including non-irradiated MOSS sensors and babyMOSS sensors subjected to different TID and NIEL irradiation levels. All tests were performed with the same set-up: a telescope composed of six ALPIDE chip layers for track reconstruction, with a MOSS chip positioned at the center as the device under test. Measurements were performed region by region, for several half units, under temperature control and monitoring, with a reverse bias of -1.2 V. Data acquisition was managed using the EUDAQ2 framework~\cite{h}, while offline data processing and track reconstruction were carried out with the Corryvreckan framework~\cite{i}. \\
Figure~\ref{fig:tb} (top) shows detection efficiency and fake-hit rate as a function of threshold for the four regions of a top half-unit in a non-irradiated MOSS sensor. The target performances, an efficiency higher than 99$\%$ and a fake-hit rate lower than 10$^{-6}$ hits/pixel/event, are indicated by the grey dashed lines. The results demonstrate consistent behavior across different regions, with similar efficiency and fake-hit rate. Additionally, a sufficient operational margin is observed, defined as the threshold range where the fake-hit rate remains below 10$^{-6}$ while maintaining a detection efficiency above 99$\%$. Figure~\ref{fig:tb} (bottom) shows detection efficiency and fake-hit rate at different threshold values for a single region of a top half unit, comparing results across various irradiation levels. Data from babyMOSS chips irradiated to ITS3 requirements are shown. A detailed performance evaluation after NIEL and TID irradiation is still in progress, including calibration and chip-to-chip variations. Preliminary results indicate that detection efficiency is only slightly affected, while TID irradiation leads to a significant increase in the fake-hit rate. However, the latest radiation load estimates for ITS3 are lower, around 400 krad and 4 $\times$ 10$^{12}$ 1 MeV n$_{\textrm{eq}}$ cm$^{-2}$,  suggesting that a sufficient operational margin remains.

\section{Conclusions}

ITS3 sensor development is progressing as planned, with key R$\&$D milestones successfully achieved. The 65 nm CMOS pixel technology has been validated, and the first large-area stitched sensors have been fabricated and are undergoing extensive characterization. Initial results demonstrate a detection efficiency above 99$\%$, with a fake hit rate below 0.1 pixel$^{-1}$ s$^{-1}$, meeting the requirements of the ITS3 for LHC Run 4. Ongoing tests focus on assessing radiation hardness, studying chip-to-chip performance variations, and improving calibration to ensure optimal operation.








\end{document}